\documentclass[superscriptaddress,reprint,amsmath,amssymb,aps,floatfix]{revtex4-1}

\usepackage{amsmath,gensymb,textcomp,bm,dcolumn,eurosym,array,tabu,multirow,nicefrac,color,subfigure,graphicx,upgreek}
\usepackage[colorlinks, linkcolor=blue, citecolor=blue, urlcolor=blue, breaklinks=true]{hyperref}

\usepackage{mathpazo,times} 

\newcommand{\ket}[1]{\lvert #1 \rangle}
\newcommand{\braket}[2]{\langle #1 \vert #2 \rangle}

\usepackage[subsectionbib]{bibunits}
\usepackage{graphicx}
\usepackage{dcolumn}
\usepackage{bm,MnSymbol}
\usepackage{units}
\usepackage{psfrag}
\usepackage{float}
\usepackage{color}

\begin{document}
\title{Quantum generative adversarial learning for simultaneous multiparameter estimation}

\author{Zichao Huang}
\affiliation{Department of Physics and Collaborative Innovation Center for Optoelectronic Semiconductors and Efficient Devices, Xiamen University, Xiamen 361005, China}
\author{Yuanyuan Chen}
\email{chenyy@xmu.edu.cn}
\affiliation{Department of Physics and Collaborative Innovation Center for Optoelectronic Semiconductors and Efficient Devices, Xiamen University, Xiamen 361005, China}
\author{Lixiang Chen}
\email{chenlx@xmu.edu.cn}
\affiliation{Department of Physics and Collaborative Innovation Center for Optoelectronic Semiconductors and Efficient Devices, Xiamen University, Xiamen 361005, China}
%


\begin{abstract}
Generative adversarial learning is currently one of the most prolific fields in artificial intelligence due to its great performance in a variety of challenging tasks such as photorealistic image and video generation. While a quantum version of generative adversarial learning has emerged that promises exponential advantages over its classical counterpart, its experimental implementation and potential applications with accessible quantum technologies remain explored little. Here, we report an experimental demonstration of quantum generative adversarial learning with the assistance of adaptive feedback that is based on stochastic gradient descent algorithm. Its performance is explored by applying this technique to the adaptive characterization of quantum dynamics and simultaneous estimation of multiple phases. These results indicate the intriguing advantages of quantum generative adversarial learning even in the presence of deleterious noise, and pave the way towards quantum-enhanced information processing applications.
\end{abstract}
\maketitle

\section{Introduction}
Generative adversarial networks (GAN) has been widely used for image and video processing, pattern recognition, secure steganography and molecule development \cite{salimans2016improved,wang2018multiscale,ledig2017photo}. In GAN algorithm, a generator tries to optimize her strategy over a number of trials, and generate statistics of data to make a discriminator unable to discriminate between the generated data and real data. Thus, the generator and the discriminator can be thought to be adversaries in a machine learning game. The endpoint of such an adversarial game is the unique Nash equilibrium, indicating that the generator finds the correct statistics of data and the discriminator unable to tell the difference \cite{heusel2017gans}. Although GAN works well in training the generator to learn the statistics of true data and leads to significant interest in industries, the requirement of huge computing power makes them to be big challenges for classical computer that is limited by Moore's law. In particular, with the rapid growth of data in volume and dimensionality, the classical GAN model is getting exponentially larger and require greatly increased computational resource \cite{sagiroglu2013big,marx2013the}. To deal with these complicated learning tasks, quantum version of GAN (QGAN) has aroused widespread research interest both in theory and experiment \cite{biamonte2017quantum,lloyd2018quantum,dallaire-demers2018quantum,zoufal2019quantum,hu2019quantum}. As quantum information processing can exhibit an exponential advantage over classical counterparts, QGAN has the potential to provide speed-up over the classical algorithms \cite{he2016deep,bishop2006pattern,duchi2011adaptive,goodfellow2014generative}. However, the experimental implementation of QGAN with accessible quantum resources in current stage remains explored little.

Here, we experimentally demonstrate a quantum version of generative adversarial network (QGAN) with the assistance of adaptive feedback that is based on stochastic gradient descent algorithm (namely the \textit{self-guided QGAN}), wherein both the processing datasets and the projective measurement are quantum from the beginning. Self-guided QGAN can work with any multilevel systems, here we use photonic qubits because they are conveniently prepared and measured at room temperature, which can also be readily extended to a quantum system with higher dimensions \cite{zoufal2019quantum,hu2019quantum}. In order to evaluate its performance, we apply this self-guided QGAN to the adaptive characterization of quantum dynamics (ACQD). While the conventional characterization methods face notorious challenges in performing the full Bell-state measurement (BSM) \cite{mohseni2006direct,mohseni2007direct,graham2013hyperentanglement,wang2007experimental} or requiring a large number of ensemble measurements \cite{kim2020universal,luchnikov2020machine,shabani2011efficient}, our QGAN algorithm uses Hong-Ou-Mandel (HOM) interference to obtain the distinguishability in a single-shot measurement that eludes the requirement of BSM \cite{hong1987measurement,nagali2009optimal}, and this distinguishability works as a realtime feedback for self-guided learning that decreases the number of experimental trials. Additionally, we also explore the application of self-guided QGAN in the simultaneous estimation of multiple phases, which is a crucial tool for quantum-enhanced sensing and imaging, and may more complex quantum computation and quantum communication protocols \cite{gorecki2022multiple,liu2021distributed,humphreys2013quantum,hong2021quantum,pezze2017optimal}.

While machine learning has been widely used for enhancing the performance of quantum state tomography (QST)\cite{prl2016experiment,rambach2021robust,ferrie2014self,hou2020experimental}, we identify four main advantages associated with our QGAN method using HOM interference. (i) It is well known that the projective measurement is still a technical challenge for frequency-encoded and time-encoded quantum states since they have the requirements of inefficient frequency shifter \cite{guo2017testing} or complicated Fresnel interferometer \cite{bulla2022non,vedovato2018postselection,marcikic2004distribution}. Conversely, HOM interference can measure the distinguishability in any degree of freedom, thus our QGAN method is available in a wider range of applications. (ii) Since HOM interference can reveal the information of incident entangled state, our QGAN method has the potential to characterize the quantum entanglement accordingly. (iii) HOM interference is determined by the distinguishable information in all degree of freedom, our QGAN method can learn the quantum state encoded in multiple degrees of freedom simultaneously, which is an essential tool for hyperentanglement-based quantum information processing and quantum microscopy \cite{ndagano2022quantum}. (iv) As a direct result of inherent stability of HOM interference measurement, our QGAN approach promises great robustness against deleterious noise and experimental imperfection, which has the potential to be widely used in the future practical applications \cite{hong1987measurement}.

Thus this scheme may provide an alternative approach for fundamental test of quantum mechanics, and shedding the light on more quantum information processing tasks that are significantly improved by quantum machine learning algorithms \cite{xia2018quantum,schuld2019quantum,cai2015entanglement}.

\begin{figure}[!t]
\centering
\includegraphics[width=\linewidth]{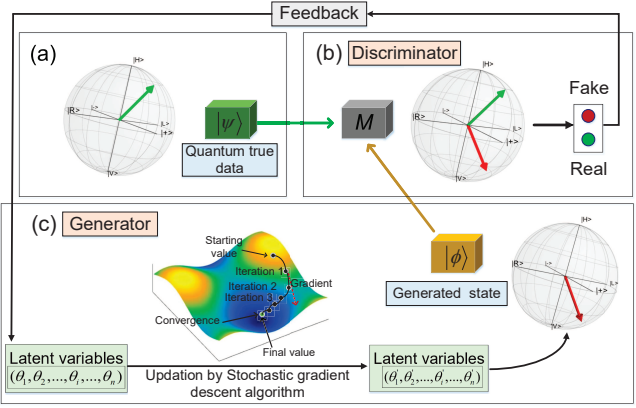}
\vspace{-7mm}
\caption{Schematic of self-guided quantum generative adversarial network, wherein the quantum states are depicted on the Bloch spheres. (a) The quantum true state described by $\ket{\psi}$ is provided by a quantum system whose internal physical structure and quantum process is not required to be known. (b) The discriminator plays a role in distinguishing the quantum true state $\ket{\psi}$ and fake state $\ket{\phi}$, which is shown as green and red Bloch vector respectively. (c) The generator takes the discriminator's measurement outcome as a real-time feedback to update the latent variables of fake state by using stochastic gradient descent algorithm.}
\label{figure_1}
\end{figure}
\section{Self-guided QGAN by using quantum interference}\label{section:2}
Figure \ref{figure_1} shows the schematic of self-guided QGAN method. The quantum true data described by $\ket{\psi}$ is provided by a quantum system, whose internal physical structure and quantum process is not required to be known. The generator can prepare arbitrary quantum data that is described by $\ket{\phi}$, whose purpose is to mimic the quantum true data with certain probability. The discriminator $M$ performs joint quantum measurement on the quantum true data and the generated fake data, and attempts to distinguish between them by analyzing the statistics of measurement results. In the self-guided QGAN, the generator can receive the information about the measurement outcomes that works as a real-time feedback, and then adaptively adjusts his strategies alternatively to fool the discriminator, until the discriminator cannot distinguish the difference between the true data and fake data any more. Thus, the generator learns the statistics of the quantum true data with great advantages in precision and accuracy.

Compared to other QGAN methods, we solve the optimization problem in the process of the generator by using stochastic gradient descent algorithms \cite{rambach2021robust,ferrie2014self}. It finds the correct quantum true data by iteratively maximizing the overlap between the quantum true data and the generated fake data using two quantum interference measurements per iteration independently. In each iteration, the algorithm measures the similarity of the quantum true data and two random fake data as $f(\ket{\phi_\pm})=|\braket{\phi_\pm}{\psi}|^2$ to approximate a gradient which heads to the true data more rapidly and then update the generated fake data. Here, the quantum data are encoded into two coherent single photons, which are routed into different input ports of the beam splitter in a HOM interferometer simultaneously \cite{hong1987measurement,chen2018polarization}. Thus the bunching probability is directly related to the photons' level of indistinguishability, namely the overlap of two incident photons as described by $|\braket{\phi_\pm}{\psi}|^2$.

In a specific iteration $k$, the algorithm chooses one random direction as $\Delta_k\in\{1,-1,i,-i\}$, and generates two projection states as $\ket{\phi_\pm}=\ket{\phi_k\pm\beta_k\Delta_k}$. Thereinto, $\beta_k=b/(k+1)^t$, where $(b,t)$ are hyperparameters for controlling the gradient estimation step size. The system then calculates a gradient $g_k $based on the similarity measurement results as
\begin{equation}\label{eq: gradient}
	g_k=\frac{f(\ket{\phi_+})-f(\ket{\phi_-})}{2\beta_k}\Delta_k.
\end{equation}
Since the visibility of HOM interference can reveal the indistinguishable level between two measured photons, it can be used directly in Eq.~\ref{eq: gradient} by replacing as
\begin{equation}
\begin{split}
f(\ket{\phi_+})-f(\ket{\phi_-})&=f(\ket{\phi_k+\beta_k\Delta_k})-f(\ket{\phi_k-\beta_k\Delta_k})\\
&\propto\frac{|\braket{\phi_k+\beta_k\Delta_k}{\psi}|^2-|\braket{\phi_k-\beta_k\Delta_k}{\psi}|^2}{|\braket{\phi_k+\beta_k\Delta_k}{\psi}|^2+|\braket{\phi_k-\beta_k\Delta_k}{\psi}|^2}\\
&\propto V_+-V_-,
\end{split}
\end{equation}
where $V_\pm$ denote the visibilities of HOM interference from the overlap measurements between the projection states $\ket{\phi_\pm}$ and the quantum true state $\ket{\psi}$ for each iteration of self-guided QGAN. Since we are interested in the gradient direction in which we get higher overlap probability rather than the actual values of the coincidence counts, the interference visibilities are all normalized for the sake of simplicity.
Then, the generated state is updated by following
\begin{equation}
\ket{\phi_{k+1}}=\ket{\phi_k+\alpha_kg_k},
\end{equation}
where $\alpha_k=a/(k+1+A)^s$, $a$, $A$, and $s$ are hyperparameters for controlling the convergence step size. This procedure is repeated for each iteration by updating $\ket{\phi_k}\rightarrow\ket{\phi_{k+1}}$ until the discriminator cannot distinguish the difference between the ture state and fake state. As shown in the inset of Fig.~\ref{figure_1}\textcolor{blue}{(c)}, the self-guided QGAN is steered by the gradient $g_k$ and will thus converge to the underlying state after a sufficient number of iterations. The values of these hyperparameters can depend on the specific system that is under investigation. While it is possible to arbitrarily optimize the fidelity by increasing the number of iterations, it can be chosen to match the required accuracy and iterative speed that typically depends on the practical applications. 
\section{Self-guided QGAN for adaptive characterization of quantum dynamics}
\begin{figure}[!t]
	\centering
	\includegraphics[width=\linewidth]{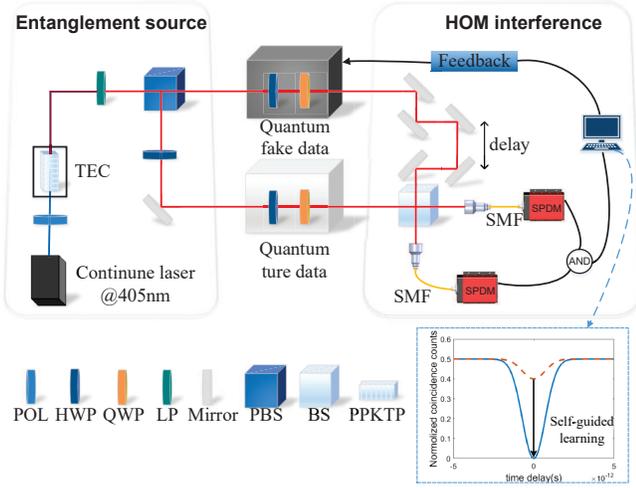}
	\vspace{-7mm}
	\caption{Experimental implementation of self-guided QGAN. POL: polarizor; HWP: half-wave plate; QWP: quarter-wave plate; TEC: temperature controller; PPKTP: periodically poled potassium titanyl phosphate nonlinear crystal; LP: longpass filter; PBS: polarizing beam splitter; BS: beam splitter; SPDM: single photon detection module; $\ket{\phi}$: quantum fake state; $\ket{\psi}$: quantum ture state. As shown in the inset, the visibility of the well-known HOM dip will gradually drop to the minimum value with the self-guided algorithm, and the measured results of HOM interference work as a real-time feedback to update the latent variables of quantum fake data $\ket{\phi}$. }
	\label{figure_2}
\end{figure}
The evolution of a quantum system can be expressed in terms of a completely positive quantum dynamical map $\varepsilon$ as $\varepsilon(\rho)=\Sigma_{m,n=0}^{d^2-1}\chi_{mn}E_m\rho E_n^{\dag}$, where $\rho$ is the incident initial state, $\{E_m\}$ are a set of operator basis elements, $\{\chi_{mn}\}$ are the matrix elements of the superoperator that encodes all the information about the dynamics. Typically, pauli-matrix rotation quantum processes can be implemented using wave plates \cite{mohseni2006direct,mohseni2007direct}. By applying a general term, an arbitrary combination of unitary matrices turns the initial state into $\ket{\psi}_t=(\alpha\ket{H}+\beta\ket{V})/\sqrt{2}$, where the complex probability amplitudes satisfied the normalization condition as $|\alpha|^2+|\beta|^2=1$. Then our self-guided QGAN method can be used to learn this quantum state with high accuracy as shown in Fig.\ \ref{figure_2}, which consequently reveals the the corresponding matrix elements of $\{\chi_{mn}\}$.

The single photons are prepared via spontaneous parametric down-conversion \cite{zhong2018photon}. In our experimental realization, a 5-mm-long ppKTP crystal is pumped with a continuous-wave pump laser, which provides type-II collinear phase matching with pump (p), signal (s), and idler (i) photons at center wavelengths of $\lambda_p\approx\unit[405]{nm}$ and $\lambda_{s,i}\approx\unit[810]{nm}$ at a temperature of $28^\degree C$. The pumping of this nonlinear crystal for SPDC emission is $\ket{H}_p\rightarrow(\ket{H}_s\ket{V}_i+\ket{V}_s\ket{H}_i)/\sqrt{2}$. These photons are guided to a polarizing beam splitter, which maps the orthogonally polarized photon pairs into two distinct spatial modes \cite{chen2020verification}. As a direct result, paired single photons that are separated in distinct spatial modes are routed to the input ports of a beam splitter. Thereinto, one of the paired photons pass through a series of wave plates, which changes the single photons' state to $\ket{\psi}_t$. Its nonclassical beating can be observed by scanning the time of arrival of one of the photons incident on this beam splitter, which constitutes a HOM interferometer. The corresponding interference fringes can be observed in the twofold coincidences between the two output ports of the beam splitter. Namely, by detecting the HOM interference, it is possible to perform a similarity measurement on the incident paired photons without any requirements of any prior knowledge. In each iteration, we route two independent single photons in quantum states $\ket{\psi}_t$ and $\ket{\phi_\pm}$ to a HOM interferometer. The detection in coincidence at two output ports reveals the overlap between them, i.e., $V_{\pm}\propto|\braket{\phi_\pm}{\psi}_t|^2$, which can be used to calculated the current gradient by following Eq.~\eqref{eq: gradient}. These measurement results would be sent back to the generator, and form the basis for updating the latent variables of generated state $\ket{\phi}_t$. Since the maximum overlap means that the coincidence detection reaches its minimum value (the well-known HOM dip), the self-guided QGAN reaches the Nash equilibrium. According to the analysis of HOM interference, this indicates that two incident quantum states are indistinguishable in all degrees of freedom, i.e., $\ket{\phi}=\ket{\psi}_t$ \cite{frederic2017high}.

\begin{figure*}[!t]
	\centering
	\includegraphics[width=\linewidth]{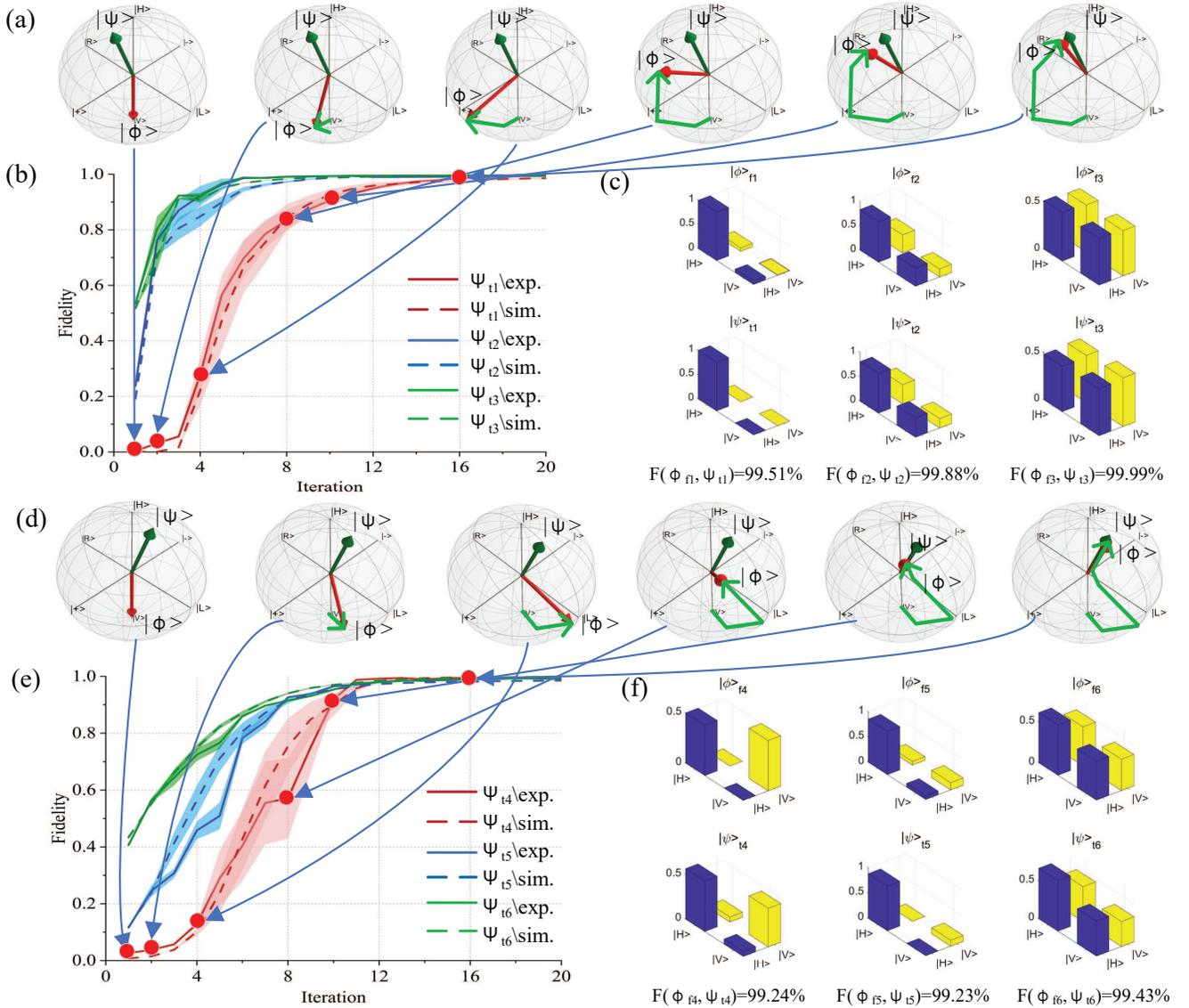}
	\vspace{-7mm}
	\caption{Tracking of the self-guided QGAN. (a-c) Experimental results for three quantum true states $\ket{\psi}_{t1,t2,t3}$ that are prepared by rotating a HWP. (a) The snapshots of the system at the particular steps at iteration 1, 2, 4, 8, 10, 16 in the Bloch sphere representation, in which the quantum true data and fake data is represented by the green and red Bloch vectors respectively. (b) The tracking of fidelity during the quantum adversarial learning process. (c) The density matrices of quantum true states $\ket{\psi}_t$ and experimentally estimated states $\ket{\phi}_f$, where an average fidelity of $99.8\%$ is achieved. (d-f) Experimental results for three quantum true states $\ket{\psi}_{t4,t5,t6}$ that are prepared by rotating a HWP and a QWP, where each panel is the counterpart of (a) to (c) respectively. The experimentally measured average fidelity reaches $99.3\%$.}
	\label{figure_3}
\end{figure*}
Figure \ref{figure_3}\textcolor{blue}{(a-c)} shows the experimental results for learning three random quantum true states $\ket{\psi}_{t1}$=$\ket{H}$, $\ket{\psi}_{t2}$=0.90$\ket{H}$+0.44$\ket{V}$, $\ket{\psi}_{t3}$ = 0.69$\ket{H}$ + 0.72$\ket{V}$, which are prepared by rotating a HWP that is relevant to a single parameter to be estimated. Additionally, a series of HWP and QWP are used to prepare three random quantum true states $\ket{\psi}_{t4}$ = 0.94$\ket{H}$ + 0.34$\ket{V}$, $\ket{\psi}_{t5}$ = 0.75$\ket{H}$ + (0.07+0.65i)$\ket{V}$, $\ket{\psi}_{t6}$ = 0.82$\ket{H}$ + (0.57+0.11i)$\ket{V}$, which are relevant to two independent parameters to be estimated, and the experimental results are depicted in Fig.~\ref{figure_3}\textcolor{blue}{(d-f)}. The initial quantum fake state is set as $\ket{\phi}_i=\ket{V}$ in both scenarios. During the self-guided QGAN method, the trajectories of fidelity are recorded [see Fig.~\ref{figure_3}\textcolor{blue}{(b,e)}] instead of characterizing the exact experimental $\ket{\phi}$. Thereinto, the solid lines and dashed lines represent the average fidelities of 5 experimental trials and 100 theoretical simulations, and the shaded bands describe the standard deviation of errors. These final experimental results are consistent with the theoretical predictions, where the slight deviations can be mainly attributed to the precision of detector and the randomness of stochastic gradient descent algorithm. The snapshots of the quantum states and measurement axis at the particular steps derived from the target states $\ket{\psi}_{t1}$ and $\ket{\psi}_{t4}$ are plotted on the Bloch sphere as shown in Fig.~\ref{figure_3}\textcolor{blue}{(a,d)}. While the generator learns from the measurement outcomes and follows stochastic gradient descent algorithm, the generated quantum fake data (shown as the red Bloch vectors) gradually converges to the quantum true data (shown as the green Bloch vectors). As a direct result, the discriminator that use HOM interference ultimately fails to discriminate $\ket{\phi}_f$ and $\ket{\psi}_t$, and correspondingly the generator achieves its goal of replicating the statistics of the quantum true data. In order to evaluate the performance of self-guided QGAN, we analyze the state fidelity to quantify the indistinguishability between the quantum true data and the generated data as $F(\phi_f,\psi_t)=tr\sqrt{\sqrt{\psi_t}\phi_f\sqrt{\psi_t}}$. As shown in Fig.~\ref{figure_3}\textcolor{blue}{(c,f)}, the average fidelity of $99.6\%$ is achieved after 20 iterations in our proof-of-principle experiment, which can be further enhanced by increasing the number of iterations and improving the experimental implementation such as higher signal-to-noise ration or higher detection efficiency. We note that our approach can reach its equilibrium without any requirement of prior knowledge about the quantum true data or the HOM interference measurement performed by the discriminator, and thus promises a double-blind quantum machine learning process just as its classical counterparts.

\begin{figure}[!t]
	\centering
	\includegraphics[width=\linewidth]{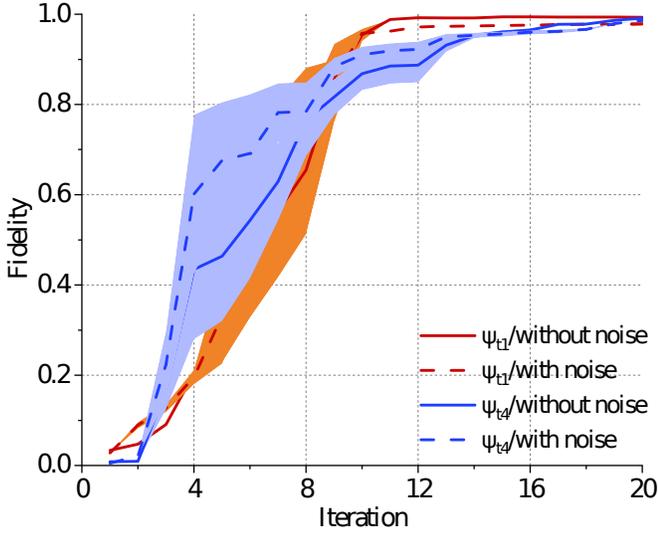}
	\caption{Performance of self-guided QGAN in the presence of environmental noise for quantum true states $\ket{\psi}_{t1}$ and $\ket{\psi}_{t4}$. The dashed lines and solid lines represent the average fidelities of 5 experimental trials with- and without-noise respectively, and the shaded bands show the standard deviation of errors. }
	\label{figure_4}
\end{figure}

The performance of self-guided QGAN in the presence of significant environmental noise is also investigated as shown in Fig.~\ref{figure_4}. We choose $\ket{\psi}_{t1}$ and $\ket{\psi}_{t4}$ as the quantum true states and introduce the significant statistical noise from a source of light emitting diode. These results show excellent performance in the presence and absence of noise, which indicates the great robustness of our approach against deleterious noise.

\section{Self-guided QGAN for multiple phase estimation}
Quantum metrology exploits quantum mechanics to enable higher precision than its classical counterpart. This has been widely explored with the estimation of optical phase shifts by means of interferometry providing the dominant paradigm \cite{chen2019hong,lyons2018attosecond}. One of the most important metrology problems to the wider research community is the microscopy and imaging \cite{ndagano2022quantum}. Phase imaging is a cornerstone of optical microscopy, typically realized using the related techniques of phase contrast and differential interference contrast imaging. Our approach maps phase imaging onto the problem of multiple phase estimation simultaneously \cite{pezze2017optimal,hong2021quantum,humphreys2013quantum,gorecki2022multiple}. We provide a strategy for the estimation of multiple phases using self-guided QGAN method, in which the multiparameter nature of the problem leads to an intrinsic benefit when exploiting quantum resources.

As shown in Fig. \ref{figure_2}, since the difference frequency of down-converted photons typically exceeds that of the pump laser, high-dimensional frequency entanglement arises quite naturally as a consequence of energy conservation, and can be written in the form of
\begin{equation}
\ket{\Psi}_\omega=\sum_k^n A_k (\ket{\omega_s^k\omega_i^k}+\ket{\omega_i^k\omega_s^k}),
\end{equation}
where $A_k$ represents the probability amplitude of the $k-th$ frequency-bin entanglement, $n$ donates the dimensions of prepared entanglement, $\omega_{s,i,p}$ represent the center frequencies of signal, idler and pump photons respectively, and they satisfy the energy conservation as $\omega_s^k+\omega_i^k=\omega_p$. By encoding the quantum true data that is described by multiple unknown phases $\{\psi_1, \psi_2, ..., \psi_n\}$, a commercial wave shaper can be readily used to modulate the relative phase shifts on the signal photon independently. Meanwhile, the generator also uses a wave shaper to encode the imitated quantum data $\{\phi_1, \phi_2, ..., \phi_n\}$ on the paired idler photons. Accordingly, the initial entangled state is transformed into
\begin{equation}
\ket{\Psi}_\omega\rightarrow\sum_k^n A_k (\ket{\omega_s^k\omega_i^k}+e^{i(\psi_k-\phi_k)}\ket{\omega_i^k\omega_s^k}).
\end{equation}
Then the discriminator uses HOM interference as a projective measurement to distinguish the quantum true data and the generated data. Specifically, in each iteration, paired photons are routed to a balanced HOM interferometer, and the coincidence counts in two opposite spatial modes are identified. The manifestation of interference fringes can be approximated by the sum of coincidence probabilities with difference frequency detunings as \cite{chen2021temporal}
\begin{equation}
P(\tau)=\sum_k^n \frac{A_k}{2}[1-cos(\psi_k-\phi_k)exp(-\sigma^2\tau^2)],
\end{equation}
where $\sigma$ is the RMS (root mean square) bandwidth of SPDC photons. For visibility and efficiency, we can set $\tau=0$ such that the $P(\tau)$ is only determined by the term of $\phi_k-\varphi_k$. Analogously, the measured interference probability is used to calculate the current gradient by following \eqref{eq: gradient}, which would be sent back to the generator for updating the latent variables and forming the basis for next iteration. Since the maximum overlap means that the coincidence detection reaches its maximum value (the well-known HOM peak), the self-guided QGAN reaches the Nash equilibrium, namely two incident quantum states are indistinguishable in all degrees of freedom, i.e., $\psi_k=\phi_k$ for all $k=1,2,...,n$ simultaneously.

\begin{figure}[!t]
\centering
\includegraphics[width=\linewidth]{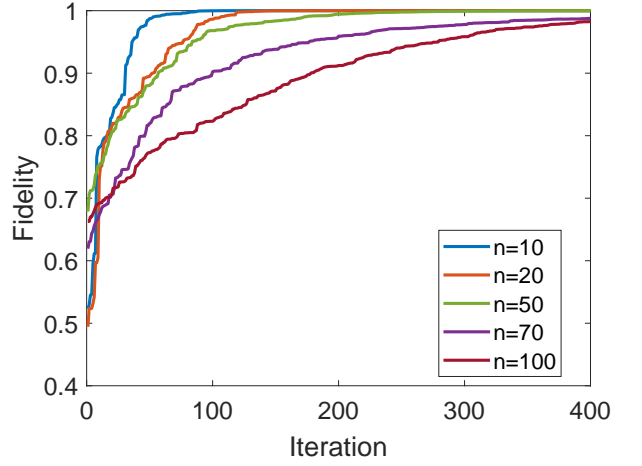}
\caption{Theoretical simulation of self-guided QGAN for simultaneous estimation of 10, 20, 50, 70 and 100 phases.}
\label{figure_5}
\end{figure}

To evaluate the performance of our self-guided QGAN for multiphase estimation simultaneously, Fig.\ \ref{figure_5} demonstrates the theoretical simulation results, wherein the estimation accuracy can reach $0.98$ for the estimation of $100$ unknown phases simultaneously. We note again that the gradient estimation step and convergence can be tuned by the corresponding parameters in the stochastic gradient descent algorithm.

\section{Discussion}
We present the concept and experiment of self-guided QGAN for simultaneous multiparameter estimation with the assistance of adaptive feedback that is based on stochastic gradient descent algorithm. Thereinto, the quantum true data and the generated fake data are all encoded using quantum superposition states, and the discriminator performs the HOM interference as the projective measurement. We also explore the robustness of self-guided QGAN against deleterious experimental and environmental noise, and show excellent stability as a direct result of quantum interference measurement.

While the self-guided QGAN approach has be used for adaptive tomography of process dynamics and simultaneous estimation of multiple phases as demonstrated in this work, it can be readily extended to analogous applications like quantum image generation \cite{huang2021experimental} and quantum state tomography \cite{rambach2021robust}. Additionally, the self-guided QGAN algorithm can be straightforwardly extended to a quantum system with higher dimensions without any requirements of modification of the experimental setup, which consequently paves the way to tackle more complicated information processing tasks. In our architecture, the bosonic mode actually provides a quantum system with infinite dimensions, the whole approach can be extended to other platform like superconducting circuit \cite{hu2019quantum}. Another possible extension of our current experiment is to explore a more complicated architecture with multiple photonic modes, wherein the two-photon HOM interferometric configuration should be replaced with multiport quantum interference \cite{zhong2018photon}. Our work may inspire more theoretical and experimental researches into the advanced quantum machine learning algorithms that provides great advantages over the classical methods.

\section*{Acknowledgements}
This work is supported by the National Natural Science Foundation of China (NSFC) (12034016, 12004318, 61975169), the Fundamental Research Funds for the Central Universities at Xiamen University (20720190057, 20720210096), the Natural Science Foundation of Fujian Province of China (2020J05004), the Natural Science Foundation of Fujian Province of China for Distinguished Young Scientists (2015J06002), and the program for New Century Excellent Talents in University of China (NCET-13-0495).

\bibliography{apssamp}

\end{document}